\documentstyle[sprocl,epsfig]{article}

\def\Journal#1#2#3#4{{#1} {\bf #2}, #3 (#4)}

\def\NPB{{\em Nucl. Phys.} B} 
\def\PLB{{\em Phys. Lett.}  B}
\def\PRL{\em Phys. Rev. Lett.} 
\def\PRD{{\em Phys. Rev.} D}
\def\ZPC{{\em Z. Phys.} C}

\def\ra{\rightarrow}
 
\def\dm21{\delta M^2_{21}}
\def\be{\begin{equation}} 
\def\ee{\end{equation}}
\def\bea{\begin{eqnarray}}
\def\eea{\end{eqnarray}}

\def\Mlq{M_{\ell \;{\rm or}\; q}}
\def\eqn#1{Eq.~(\ref{eq#1})}  
\def\bra#1{\langle#1|} 
\def\ket#1{|#1\rangle}

\def\gsim{\;\raisebox{-.6ex}{$\stackrel{>}{\sim}$}\;}

\def\decayarrow{\kern0.2em\hbox{$\raise1.08ex\hbox{\big|}\kern-0.5em
\longrightarrow$}}

\begin{document}

\title{NEUTRINO MASS, MIXING, AND OSCILLATION\thanks{} }

\author{Boris Kayser}

\address{National Science Foundation, 4201 Wilson Blvd. \\ Arlington, VA
 22230, USA\\E-mail: bkayser@nsf.gov}

\maketitle\abstracts{Do neutrinos have nonzero masses? If they do, then these masses are very tiny, and can be sought only in very sensitive experiments. The most sensitive of these search for {\em neutrino oscillation}, a quantum interference effect which requires neutrino mass and leptonic mixing. In these lectures, we explain what leptonic mixing is, and then develop the physics of neutrino oscillation, including the general formalism and its application to special cases of practical interest. We also see how neutrino oscillation is affected by the passage of the oscillating neutrinos through matter.}

\section{Introduction}

\footnotetext{To appear in the Proceedings of TASI 2000, the Theoretical Advanced Study Institute in elementary particle physics held in Boulder, Colorado in June 2000.}

We humans, and the everyday objects around us, are made of nucleons and electrons, so these particles are the ones most familiar to us. However, for every nucleon or electron, the universe as a whole contains around a billion neutrinos. In addition, each person on earth is bombarded by $\sim 10^{14}$ neutrinos, coming from the sun, every second. Clearly, neutrinos are abundant. Thus, it would be very nice to know something about them.

One of the most basic questions we can ask about neutrinos is: Do they have nonzero masses? Until recently, there was no hard evidence that they do. It was known that, at the heaviest, they are very light compared to the quarks and charged leptons. But for a long time, the natural theoretical prejudice has been that the neutrinos are {\em not} massless. One reason for this prejudice is that in essentially any of the grand unified theories that unify the strong, electromagnetic, and weak interactions, a given neutrino $\nu$ belongs to a large multiplet $F$, together with at least one charged lepton $\ell$, one positively-charged quark $q^+$, and one negatively-charged quark $q^-$:
\be
F = \left[ \begin{array}{c}
			\nu \\ \ell \\ q^+ \\ q^-
			\end{array}  \right] \;\; .
\label{eq1}
\ee
The neutrino $\nu$ is related by a symmetry operation to the other members of $F$, all of which have nonzero masses. Thus, it would be peculiar if $\nu$ did not have a nonzero mass as well.

If neutrinos do have nonzero masses, we must understand why they are nevertheless so light. Perhaps the most appealing explanation of their lightness is the ``see-saw mechanism''.\cite{b1} To understand how this mechanism works, let us note that, unlike charged particles, neutrinos may be their own antiparticles. If a neutrino is identical to its antiparticle, then it consists of just two mass-degenerate states: one with spin up, and one with spin down. Such a neutrino is referred to as a Majorana neutrino. 
In contrast, if a neutrino is distinct from its antiparticle, then it plus its antiparticle form a complex consisting of four mass-degenerate states: the spin up and spin down neutrino, plus the spin up and spin down antineutrino. This collection of four states is called a Dirac neutrino. In the see-saw mechanism, a four-state Dirac neutrino ${\cal N}^D$ of mass $M_D$ gets split by ``Majorana mass terms'' into a pair of two-state Majorana neutrinos. One of these Majorana neutrinos, $\nu^M$, has a small mass $M_\nu$ and is identified as one of the observed light neutrinos. 
The other, $N^M$, has a large mass $M_N$ characteristic of some high mass scale where new physics beyond the range of current particle accelerators, and responsible for neutrino mass, resides. Thus, $N^M$ has not been observed. The character of the breakup of ${\cal N}^D$ into $\nu^M$ and $N^M$ is such that $M_\nu M_N \cong M^2_D$. 
It is reasonable to expect that $M_D$, the mass of the Dirac particle ${\cal N}^D$, is of the order of $M_{\ell \;{\rm or}\; q}$, the mass of a typical charged lepton $\ell$ or quark $q$, since the latter are Dirac particles too. Then $M_\nu M_N \cong \Mlq^2$. With $\Mlq$ a typical charged lepton or quark mass, and $M_N$ very big, this ``see-saw relation'' explains why $M_\nu$ is very tiny. Note that the see-saw mechanism predicts that each light neutrino $\nu^M$ is a two-state Majorana neutrino, identical to its antineutrino.

\section{Neutrino Oscillation}

To find out whether neutrinos really do have nonzero masses, we need an experimental approach which can detect these masses even if they are very small. The most sensitive approach is the search for neutrino oscillation. Neutrino oscillation is a quantum interference phenomenon in which small splittings between the masses of different neutrinos can lead to large, measurable phase differences between interfering quantum-mechanical amplitudes.

To explain the physics of neutrino oscillation, we must first discuss leptonic ``flavor''. Suppose a neutrino $\nu$ is born in the $W$-boson decay
\be
W^+ \ra \ell^+_\alpha + \nu \; .
\label{eq2}
\ee
Here, $\alpha = e, \mu$ or $\tau$, and $\ell_\alpha^+$ is one of the positively charged leptons: $\ell^+_e \equiv e^+$, $\ell^+_\mu \equiv \mu^+$, and $\ell^+_\tau \equiv \tau^+$. Suppose that, without having time to change its character, the neutrino $\nu$ interacts in a detector immediately after its birth in the decay (\ref{eq2}), and produces a new charged lepton $\ell^-_\beta$ via the reaction $\nu$ + {\it target} $\ra \ell^-_\beta$ + {\it recoils}. It is found that the ``flavor'' $\beta$ of this new charged lepton is always the same as the flavor $\alpha$ of the charged lepton with which $\nu$ was born. 
It follows that the neutrinos produced by the $W$ decays (\ref{eq2}) to charged leptons of different flavors must be different objects. We take this fact into account by writing these decays more accurately as
\be
W^+ \ra \ell^+_\alpha + \nu_\alpha \; ; \; \alpha = e,\mu,\tau \;.
\label{eq3}
\ee
The neutrino $\nu_\alpha$, called the neutrino of flavor $\alpha$, is by definition the neutrino produced in leptonic $W$ decay in association with the charged lepton of flavor $\alpha$. As we have said, when $\nu_\alpha$ interacts to create a charged lepton, the latter lepton is always $\ell_\alpha$. 
In neutrino oscillation, a neutrino born in association with a charged lepton $\ell_\alpha$ of flavor $\alpha$ then travels for some time during which it can alter its character. Finally, it interacts to produce a second charged lepton $\ell_\beta$ with a flavor $\beta$ {\em different} from the flavor $\alpha$ of the charged lepton with which the neutrino was born. For example, suppose a neutrino is born with a muon in the pion decay $\pi^+ \ra$ Virtual $W^+ \ra \mu^+ + \nu_\mu$. Suppose further that after traveling down a neutrino beamline, this same neutrino interacts in a detector and produces, not another muon, but a $\tau^-$. At birth, the neutrino was a $\nu_\mu$. But by the time it interacted in the detector, it had turned into a $\nu_\tau$. One describes this metamorphosis by saying the neutrino oscillated from a $\nu_\mu$ into a $\nu_\tau$. 
As we will see, the probability for it to change its flavor does indeed oscillate with the distance it travels before interacting. As we will also see, the oscillation in vacuum of a neutrino between different flavors requires neutrino mass.

\begin{figure}[t]
\vskip 0.5cm
\begin{center}
\epsfig{figure=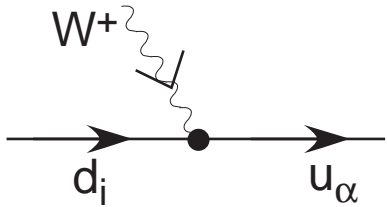} 
\end{center}
\caption{Absorption of a $W$ boson by a quark.  
\label{f1}}
\end{figure}

To see how neutrino mass can lead to neutrino oscillation, let us briefly recall the weak interactions of quarks. As we all know, there are three quarks---the $u$ (up), $c$ (charm), and $t$ (top) quarks---which carry a positive electric charge $Q = +2/3$. In addition, there are three quarks---the $d$ (down), $s$ (strange), and $b$ (bottom) quarks---which carry a negative electric charge $Q = -1/3$. Each of these six quarks is a particle of definite mass. As we know, the quarks are arranged into three families or generations, each of which contains one positive quark and one negative quark:
\begin{eqnarray*}
\begin{array}{lccc}
	Family:	&  1  	&  2	&  3  \\ [.15pc]
	Quarks:	& \left( \begin{array}{c} u \\ d \end{array} \right)
			& \left( \begin{array}{c} c \\ s \end{array} \right)
			& \left( \begin{array}{c} t \\ b \end{array} \right)
\end{array}
\end{eqnarray*}
However, we know experimentally that under the weak interaction, any of the negative quarks, $d$, $s$, or $b$, can absorb a positively-charged $W$ boson and turn into any of the positive quarks, $u$, $c$, or $t$. This is illustrated in Fig.~(\ref{f1}). 
There, $d_i (i = d, s, b)$ is one of the down-type (negative) quarks. That is, $d_d \equiv d$ is the down quark, $d_s \equiv s$ is the strange quark, and so on. Similarly, $u_\alpha (\alpha = u, c, t)$ is one of the up-type (positive) quarks, with $u_c \equiv c$ being the charm quark, and so on.

In the S(tandard) M(odel) of the electroweak interactions,\cite{b2} the $W$-quark couplings depicted in Fig.~(\ref{f1}) are described by the Lagrangian density
\be
{\cal L}_{udW} = 
	-\frac{\;g}{\sqrt2} \sum_{\stackrel{\alpha=u,c,t}{i=d,s,b}}
	\overline{u_{L\alpha}} \gamma^\lambda V_{\alpha i} d_{Li} 
	W^+_\lambda + {\rm h.c.} \;\; .
\label{eq4}
\ee
Here, the subscript $L$ denotes left-handed chiral projection. For instance, $d_{Ls} \equiv \frac{1}{2}(1-\gamma_5)d_s$ is the left-handed strange-quark field. The constant $g$ is the semiweak coupling constant, and $V$ is a $3\times 3$ matrix known as the quark mixing matrix. In the SM, $V$ is unitary, because it is basically the matrix for the transformation from one basis of quantum states to another.

\begin{figure}[t]
\vskip 0.5cm
\begin{center}
\epsfig{figure=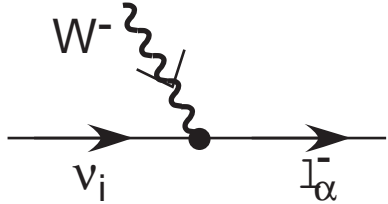} 
\end{center}
\caption{Absorption of a $W$ boson by a neutrino.  
\label{f2}}
\end{figure}

The SM interaction (\ref{eq4}) is very well confirmed experimentally. With this established behavior of quarks in mind, let us now return to the leptons. Like the quarks, the charged leptons $e, \mu$, and $\tau$ are particles of definite mass. However, if leptons behave as quarks do, then the neutrinos $\nu_e, \nu_\mu$, and $\nu_\tau$ of definite flavor are {\em not} particles of definite mass. Let us call the neutrinos which do have definite masses $\nu_i, \; i=1,2,\ldots,N$. As far as we know, the number of $\nu_i,\; N$, may exceed the number of charged leptons, three. 
Now, as we recall, the neutrino $\nu_\alpha$ of definite flavor $\alpha$ is the neutrino state that accompanies the definite-mass charged lepton $\ell_\alpha$ in the decay $W \ra \ell_\alpha + \nu_\alpha$. If leptons behave as quarks do, this neutrino state must be a superposition of the neutrinos $\nu_i$ of definite mass. 
To see this, we first note that just as any negative quark $d_i$ of definite mass can absorb a $W^+$ and turn into any positive quark $u_\alpha$ of definite mass, so it must be possible for any neutrino 
$\nu_i$ of definite mass to absorb a $W^-$ and turn into any charged lepton $\ell^-_\alpha$ of definite mass. This absorption is illustrated in Fig.~(\ref{f2}). We expect that in analogy with \eqn{4} for the $W$-quark couplings, the SM interaction that describes the $W$-lepton couplings is 
\be
{\cal L}_{\ell\nu W} = 
	-\frac{\;g}{\sqrt2}
	\sum_{\stackrel{\alpha=e,\mu,\tau}{i=1,\ldots,N}}
	\overline{\ell_{L\alpha}} \gamma^\lambda U_{\alpha i} \nu_{Li} 
	W^-_\lambda  -\frac{\;g}{\sqrt2}
	\sum_{\stackrel{\alpha}{i}}
	\overline{\nu_{Li}} \gamma^\lambda U_{i\alpha }^\dagger
	 \ell_{L\alpha} W^+_\lambda\; .
\label{eq10}
\ee
Here, $U$ is an $N\times N$ unitary matrix which is the leptonic analogue of the quark mixing matrix $V$. The matrix $U$ is referred to as the leptonic mixing matrix.\cite{b3} If $N>3$, then only the top 3 rows of $U$ enter in the $W$-lepton interaction, \eqn{10}.

The leptonic decays of the $W^+$ are governed by the second term of ${\cal L}_{\ell\nu W}$, \eqn{10}. From this term, we see that when  $W^+ \ra \ell_\alpha^+ +$ ``$\nu_\alpha$'', the neutrino state $\ket{\nu_\alpha}$ produced in association with the specific definite-mass charged lepton $\ell^+_\alpha$ is
\be
\ket{\nu_\alpha} = \sum_i U^*_{\alpha i}\, \ket{\nu_i} \; .
\label{eq11}
\ee
That is, the ``flavor-$\alpha$'' neutrino $\ket{\nu_\alpha}$ produced together with $\ell^+_\alpha$ is a coherent superposition of the mass-eigenstate neutrinos $\ket{\nu_i}$, with coefficients which are elements of the leptonic mixing matrix.

What if $N$ is bigger than three? Suppose, for example, that $N=4$. Then, with the elements of the bottom row of $U$, $U_{{\rm lastrow},i}$, we can construct a neutrino state
\be
\ket{\nu_s} \equiv \sum_i U^*_{{\rm lastrow},i}\, \ket{\nu_i}
\label{eq12}
\ee
which does not couple to any of the 3 charged leptons. This state is called a ``sterile'' neutrino, which just means that it does not participate in the SM weak interactions. It may, however, participate in other interactions beyond the SM whose effects at present-day energies are too feeble to have been observed.

\begin{figure}[t]
\vskip 0.5cm
\begin{center}
\epsfig{figure=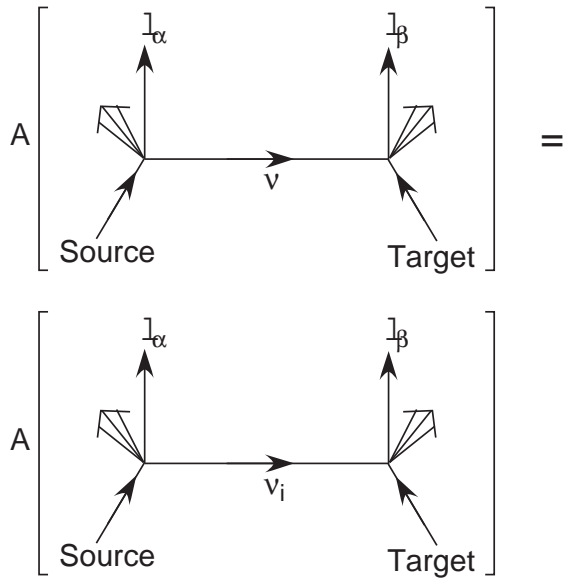,height=3.0in} 
\end{center}
\caption{Creation of a neutrino with a charged lepton of flavor $\alpha$ followed by the interaction of this neutrino to produce a charged lepton of flavor $\beta$. The ``Source'' is the particle whose decay creates the neutrino, $\ell_\alpha$, and other, unlabelled, particles. The ``Target'' is the particle struck by the neutrino to produce $\ell_\beta$ and other, unlabelled, particles. ``A'' denotes an amplitude.
\label{f3}}
\end{figure}

Owing to the leptonic mixing described by \eqn{10}, when the charged lepton of flavor $\alpha$ is created, the accompanying neutrino can be any of the $\nu_i$. Furthermore, if this $\nu_i$ later interacts with some target, it can produce a charged lepton $\ell_\beta$ of any flavor $\beta$. In such a sequence of events, the neutrino itself is an unseen intermediate state. Thus, as shown in Fig.~(\ref{f3}), the amplitude for a neutrino to be born with charged lepton $\ell_\alpha$, and then to interact and produce charged lepton $\ell_\beta$, is a coherent sum over the contributions of all the unseen mass eigenstates $\nu_i$.

The birth of a neutrino with charged lepton $\ell_\alpha^{(+)}$ and its subsequent interaction to produce charged lepton $\ell_\beta^{(-)}$ is usually described as the oscillation $\nu_\alpha \ra \nu_\beta$ of a neutrino of flavor $\alpha$ into one of flavor $\beta$ (see earlier discussion). Using ``A'' to denote an amplitude, we see from Fig.~(\ref{f3}) that
\bea
	A(\nu_\alpha \ra \nu_\beta) &=& \sum_i 
	[A(\mbox{neutrino born with }\ell^+_\alpha \mbox{ is } \nu_i) 
	\times 	\nonumber \\
	& & A(\nu_i \mbox{ propagates }) \,
	A(\mbox{when } \nu_i \mbox{ interacts it makes } \ell^-_\beta)]\;.
\label{eq13}
\eea
From ${\cal L}_{\ell\nu W}$, \eqn{10}, we find that apart from irrelevant factors,
\be
	A(\mbox{neutrino born with }\ell^+_\alpha \mbox{ is } \nu_i) = 
	U^*_{\alpha i} \; .
\label{eq14}
\ee
Similarly,
\be
	A(\mbox{when } \nu_i \mbox{ interacts it makes } \ell^-_\beta) = 
	U_{\beta i} \; .
\label{eq15}
\ee
To find the amplitude A($\nu_i$ propagates), we note that in the rest frame of $\nu_i$, where the proper time is $\tau_i$,\cite{b4}, Schr\"{o}dinger's equation states that
\be
	i\frac{\partial}{\partial\tau_i} \,\ket{\nu_i(\tau_i)} = 
	M_i \,\ket{\nu_i(\tau_i)} \; .
\label{eq16}
\ee
Here, $M_i$ is the mass of $\nu_i$. From \eqn{16},
\be
	\ket{\nu_i(\tau_i)} = e^{-i M_i\tau_i} \,\ket{\nu_i(0)} \; .
\label{eq17}
\ee
Now, for propagation over a proper time interval $\tau_i$, A($\nu_i$ propagates) is just the amplitude for finding the original state $\ket{\nu_i(0)}$ in the time-evolved $\ket{\nu_i(\tau_i)}$. That is,
\be
	A(\nu_i\mbox{ propagates})=\bra{\nu_i(0)}\nu_i(\tau_i)\rangle = 
	e^{-i M_i\tau_i} \; .
\label{eq18}
\ee
In terms of the time $t$ and position $L$ in the laboratory frame, the Lorentz-invariant phase factor $\exp{(-i M_i\tau_i)}$ is
\be
	e^{-i(E_it - p_iL)} \; .
\label{eq19}
\ee
Here, $E_i$ and $p_i$ are, respectively, the energy and momentum of $\nu_i$ in the laboratory frame. In practice, our neutrino will be highly relativistic, so if it was born at $(t,L) = (0,0)$, we will be interested in evaluating the phase factor (\ref{eq19}) where $t \approx L$, where it becomes
\be
	e^{-i(E_i - p_i)L} \; .
\label{eq20}
\ee

Suppose that the neutrino created with $\ell_\alpha$ is produced with a definite momentum $p$, regardless of which $\nu_i$ it happens to be. Then, if it is the particular mass eigenstate $\nu_i$, it has total energy
\be
	E_i = \sqrt{p^2+M^2_i} \approx p + \frac{M^2_i}{2p} \; ,
\label{eq21}
\ee
assuming that all the masses $M_i$ are much smaller than $p$. From (\ref{eq20}), we then find that
\be
	A(\nu_i \mbox{ propagates}) \approx e^{-i \frac{M^2_i}{2p} L} \; .
\label{eq22}
\ee
Alternatively, suppose that our neutrino is produced with a definite energy $E$, regardless of which $\nu_i$ it happens to be.\cite{b5} Then, if it is the particular mass eigenstate $\nu_i$, it has momentum
\be
	p_i = \sqrt{E^2-M^2_i} \approx E - \frac{M^2_i}{2E} \; .
\label{eq23}
\ee
From (\ref{eq20}), we then find that
\be
	A(\nu_i \mbox{ propagates}) \approx e^{-i \frac{M^2_i}{2E} L} \; .
\label{eq24}
\ee
Since highly relativistic neutrinos have $E \approx p$, the propagation amplitudes given by Eqs.~(\ref{eq22}) and (\ref{eq24}) are approximately equal. Thus, it doesn't matter whether our neutrino is created with definite momentum or definite energy.

Collecting the various factors that appear in \eqn{13}, we conclude that the amplitude $A(\nu_\alpha \ra \nu_\beta)$ for a neutrino of energy $E$ to oscillate from a $\nu_\alpha$ to a $\nu_\beta$ while traveling a distance $L$ is given by 
\be
A(\nu_\alpha \ra \nu_\beta) = \sum_i U^*_{\alpha i} e^{-i M_i^2 \frac{L}{2E}} U_{\beta i} \; .
\label{eq25}
\ee
The probability $P(\nu_\alpha \ra \nu_\beta)$ for this oscillation is then given by
\bea
P(\nu_\alpha \ra \nu_\beta) & = & 
	|A(\nu_\alpha \ra \nu_\beta)|^2 \nonumber \\
	& = & \delta_{\alpha\beta} - 4\sum_{i>j} \Re (U^*_{\alpha i} U_{\beta i} U_{\alpha j} U^*_{\beta j}) \sin^2 (\delta M^2_{ij} \frac{L}{4E}) \nonumber \\
	& & \hspace{.25in} + 2\sum_{i>j} \Im (U^*_{\alpha i} U_{\beta i} U_{\alpha j} U^*_{\beta j}) \sin (\delta M^2_{ij} \frac{L}{2E}) \; .
\label{eq26}
\eea
Here, $\delta M^2_{ij} \equiv M^2_i - M^2_j$, and in calculating $|A(\nu_\alpha \ra \nu_\beta)|^2$, we have used the unitarity constraint
\be
	\sum_i U^*_{\alpha i} U_{\beta i} = \delta_{\alpha\beta} \; .
\label{eq27}
\ee

Some general comments are in order:
\begin{enumerate}
	\item From \eqn{26} for $P(\nu_\alpha \ra \nu_\beta)$, we see that if all neutrino masses vanish, then $P(\nu_\alpha \ra \nu_\beta) = \delta_{\alpha\beta}$, and there is no oscillation from one flavor to another. Neutrino flavor oscillation requires neutrino mass.
	\item The probability $P(\nu_\alpha \ra \nu_\beta)$ oscillates as a function of $L/E$. This is why the phenomenon we are discussing is called ``neutrino oscillation".
	\item From \eqn{25} for $A(\nu_\alpha \ra \nu_\beta)$, we see that the $L/E$ dependence of neutrino oscillation arises from interferences between the contributions of the different mass eigenstates $\nu_i$. We also see that the phase of the $\nu_i$ contribution is proportional to $M_i^2$. 
Thus, the interferences can give us information on neutrino masses. However, since these interferences can only reveal the {\em relative} phases of the interfering amplitudes, experiments on neutrino oscillation can only determine the splittings $\delta M^2_{ij} \equiv M^2_i - M^2_j$, and not the underlying individual neutrino masses. This fact is made perfectly clear by \eqn{26} for $P(\nu_\alpha \ra \nu_\beta)$.
	\item With the so-far omitted factors of $\hbar$ and $c$ inserted,
\be\
	\delta M^2_{ij} \frac{L}{4E} = 1.27 \,\delta M^2_{ij} (\mbox{eV}^2) \frac{L\mbox{(km)}}{E\mbox{(GeV)}} \; .
\label{eq28}
\ee
Thus, from \eqn{26} for the oscillation probability $P(\nu_\alpha \ra \nu_\beta)$, we see that an oscillation experiment characterized by a given value of $L$(km) / $E$(GeV) is sensitive to mass splittings obeying
\be
	\delta M^2_{ij} (\mbox{eV}^2) \gsim 
	\left[ \frac{L\mbox{(km)}}{E\mbox{(GeV)}} \right]^{-1} \; .
\label{eq29}
\ee
To be sensitive to tiny $\delta M^2_{ij}$, an experiment must have large $L/E$. In Table~\ref{t1}, we indicate the $\delta M^2$ reach implied by \eqn{29} for experiments working with neutrinos produced in various ways.
\begin{table}[t]
\caption{The approximate reach in $\delta M^2$ of experiments studying various types of neutrinos. Often, an experiment covers a range in $L$ and a range in $E$. To construct the table, we have used typical values of these quantities.
\label{t1}}
\begin{center}
\begin{tabular}{|ccccc|}
\hline
\begin{minipage}{1.1in} \begin{center}
	\raisebox{0ex}[2ex][0ex]{Neutrinos} \\ [-2pt] (Baseline)
\end{center} \end{minipage}
	& {$L$(km)}		& {$E$(GeV)}	
	& \raisebox{0ex}[4ex][0ex]{$\frac{{\textstyle L}\mbox{(km)}}{{\textstyle E}\mbox{(GeV)}}$} 
	& \begin{minipage}{1.1in} \begin{center}	
		\raisebox{0ex}[2ex][0ex]{$\delta M^2_{ij}$(eV$^2$)} \\ [-2pt] Reach
	  \end{center} \end{minipage} \\ [8 pt]
\hline 
\begin{minipage}{1.1in} \begin{center}
	\raisebox{0ex}[3ex][0ex]{Accelerator} \\ [-2pt] (Short Baseline)
\end{center} \end{minipage}
				& $1$	& $1$		& $1$	& $1$ \\ [10 pt]
\begin{minipage}{1.2in} \begin{center}
	Reactor \\ [-2pt] (Medium Baseline)
\end{center} \end{minipage}
				& $1$		& $10^{-3}$	& $10^3$	& $10^{-3}$ \\  [8 pt]
\begin{minipage}{1.1in} \begin{center}
	Accelerator \\ [-2pt] (Long Baseline)
\end{center} \end{minipage}
				& $10^3$	& $10$		& $10^2$	& $10^{-2}$ \\ [8 pt]
Atmospheric		& $10^4$	& $1$		& $10^4$	& $10^{-4}$ \\ [8 pt]
Solar			& $10^8$	& $10^{-3}$	& $10^{11}$	& $10^{-11}$ \\ [5 pt]
\hline
\end{tabular}
\end{center}
\end{table}
	\item There are basically two kinds of oscillation experiments: {\em appearance} experiments, and {\em disappearance} experiments.

In an appearance experiment, one looks for the {\em appearance} in the neutrino beam of neutrinos bearing a flavor not present in the beam initially. For example, imagine that a beam of neutrinos is produced by the decays of charged pions. Such a beam consists almost entirely of muon neutrinos and contains no tau neutrinos. One can then look for the appearance of tau neutrinos, made by oscillation of the muon neutrinos, in this beam.

In a disappearance experiment, one looks for the {\em disappearance} of some fraction of the neutrinos bearing a flavor which {\em is} present in the beam initially. For example, imagine again that a beam of neutrinos is produced by the decays of charged pions, so that almost all the neutrinos in the beam are muon neutrinos, $\nu_\mu$. If one knows the $\nu_\mu$ flux that is produced initially, one can look to see whether some of this initial $\nu_\mu$ flux disappears after the beam has traveled some distance, and the muon neutrinos have had a chance to oscillate into other flavors.
	\item Even though neutrinos can change flavor through oscillation, the total flux of neutrinos in a beam will be conserved so long as $U$ is unitary. To see this, note that
\bea
\sum_\beta P(\nu_\alpha \ra \nu_\beta) & = & 
	\sum_\beta |A(\nu_\alpha \ra \nu_\beta)|^2 \nonumber \\
	& = & \sum_\beta 
	(\sum_i U^*_{\alpha i} U_{\beta i} e^{-i M_i^2 \frac{L}{2E}}) 
	(\sum_j U_{\alpha j} U^*_{\beta j} e^{+i M_j^2 \frac{L}{2E}}) \nonumber \\
	& = & \sum_{i,j} U^*_{\alpha i} U_{\alpha j} e^{-i \delta M^2_{ij} \frac{L}{2E}} 
	\sum_\beta U_{\beta i} U^*_{\beta j} \nonumber \\
	& = &  \sum_i |U_{\alpha i}|^2 = 1  \; .
\label{eq30}
\eea
Here, we have used \eqn{25} for the amplitude $A(\nu_\alpha \ra \nu_\beta)$ and the unitarity relations $\sum_\beta U_{\beta i} U^*_{\beta j} = \delta_{ij}$ and $\sum_i |U_{\alpha i}|^2 = 1$. The result $\sum_\beta P(\nu_\alpha \ra \nu_\beta)=1$ means that if one starts with a certain number $n$ of neutrinos of flavor $\alpha$, then after oscillation the number of neutrinos that have oscillated away into new flavors $\beta \neq \alpha$, plus the number that have retained the original flavor $\alpha$, is still $n$. 
Note, however, that some of the new flavors that get populated by the oscillation might be sterile. If they are indeed sterile, then the number of ``active'' neutrinos (i.e., neutrinos that participate in the SM weak interactions) remaining after oscillation will be less than $n$.
\end{enumerate}

\subsection{Special Cases}
Let us now apply the general formalism for neutrino oscillation in vacuum to several special cases of practical interest.

The simplest special case of all is two-neutrino oscillation. This occurs when the SM weak interaction, \eqn{10}, couples two charged leptons (say, $e$ and $\mu$) to just two neutrinos of definite mass, $\nu_1$ and $\nu_2$, and only negligibly to any other neutrinos of definite mass. It is then easily shown that the $2 \times 2$ submatrix
\be
\hat{U} = \left[ \begin{array}{cc}
		U_{e 1} & U_{e 2} \\ U_{\mu 1} & U_{\mu 2}
		\end{array} \right]
\label{eq31}
\ee
of the mixing matrix $U$ must be unitary all by itself. This means that the definite-flavor neutrinos $\nu_e$ and $\nu_\mu$ are composed exclusively of the mass eigenstates $\nu_1$ and $\nu_2$, and do not mix with neutrinos of any other flavor.

From \eqn{25} for the oscillation amplitude, we have for this two-neutrino case
\be
e^{+i M_1^2 \frac{L}{2E}} A(\nu_e \ra \nu_\mu) = U^*_{e 1} U_{\mu 1} + U^*_{e 2} U_{\mu 2} e^{-i \dm21 \frac{L}{2E}} \; .
\label{eq32}
\ee
Since $\hat{U}$ is unitary, $U^*_{e 1} U_{\mu 1} + U^*_{e 2} U_{\mu 2} = 0$, so \eqn{32} may be rewritten as
\bea
e^{+i M_1^2 \frac{L}{2E}} A(\nu_e \ra \nu_\mu) & = & 
	-U^*_{e 2} U_{\mu 2} (1 - e^{-i \dm21 \frac{L}{2E}}) \nonumber \\
	& = & -2i \, e^{-i \dm21 \frac{L}{4E}} 
	U^*_{e 2} U_{\mu 2} \sin(\dm21 \frac{L}{4E}) \; .
\label{eq33}
\eea
Squaring this result, using \eqn{28} to take account of the requisite factors of $\hbar$ and $c$, we find that the probability $P(\nu_e \ra \nu_\mu)$ for a $\nu_e$ to oscillate into a $\nu_\mu$ is given by
\be
P(\nu_e \ra \nu_\mu) = 4|U_{e 2}|^2 |U_{\mu 2}|^2 
	\sin^2 (1.27 \,\delta M^2 (\mbox{eV}^2) \frac{L\mbox{(km)}}{E\mbox{(GeV)}}) \; .
\label{eq34}
\ee
Here, we have introduced the abbreviation $\dm21 \equiv \delta M^2$. From the $e \leftrightarrow \mu$ symmetry of the right-hand side of \eqn{34}, it is obvious that
\be
P(\nu_\mu \ra \nu_e) = P(\nu_e \ra \nu_\mu) \; .
\label{eq35}
\ee

The probability $P(\nu_\alpha \ra \nu_\alpha)$ that a neutrino $\nu_\alpha$ of flavor $\alpha = e$ or $\mu$ retains its original flavor is given by
\bea
P(\nu_\alpha \ra \nu_\alpha) & = & 1 - P(\nu_\alpha \ra \nu_{\beta \neq \alpha}) \nonumber \\
& = & 1 - 4|U_{\alpha 2}|^2 (1-|U_{\alpha 2}|^2) 
	\sin^2 (1.27 \,\delta M^2 (\mbox{eV}^2) \frac{L\mbox{(km)}}{E\mbox{(GeV)}}) \; .
\label{eq36}
\eea
To obtain this expression, we have used the conservation of probability, \eqn{30}, the ``off-diagonal'' oscillation probability, Eqs.~(\ref{eq34}) and (\ref{eq35}), and the unitarity relation $|U_{e 2}|^2 + |U_{\mu 2}|^2 = 1$.

The unitarity of $\hat{U}$, \eqn{31}, implies that it can be written in the form
\be
\hat{U} = \left[ \begin{array}{c@{\hspace{.15in}}c}
		e^{i\varphi_1}\cos\theta & e^{i\varphi_2}\sin\theta \\ 
		-e^{i(\varphi_1+\varphi_3)}\sin\theta &   e^{i(\varphi_2+\varphi_3)}\cos\theta
		\end{array} \right] \;\; .
\label{eq37}
\ee
Here, $\theta$ is an angle referred to as the leptonic mixing angle and $\varphi_{1,2,3}$ are phases. From \eqn{37}, $4|U_{e 2}|^2 |U_{\mu 2}|^2 = \sin^2 2\theta$, so that $P(\nu_e \ra \nu_\mu)$, \eqn{34}, takes the form
\be
P(\nu_e \ra \nu_\mu) = P(\nu_\mu \ra \nu_e) = \sin^2 2\theta \sin^2 (1.27 \,\delta M^2(\mbox{eV}^2) \frac{L\mbox{(km)}}{E\mbox{(GeV)}})\; .
\label{eq38}
\ee
This is the most-commonly quoted form of the two-neutrino oscillation probability.

A second special case which may prove to be very relevant to the real world is a three-neutrino scenario in which two of the neutrino mass eigenstates are nearly degenerate. That is, the neutrino (Mass)$^2$ spectrum is as in Fig.~({\ref{f4}), where
\be
	|\dm21| \equiv \delta M^2_{\rm Small} \ll 
	|\delta M^2_{31}| \cong |\delta M^2_{32}| \equiv 
	\delta M^2_{\rm Big} \; .
\label{eq39}
\ee
All three of the charged leptons, $e, \mu$, and $\tau$, are coupled by the SM weak interaction, \eqn{10}, to the neutrinos $\nu_{1,2,3}$.
\begin{figure}[t]
\vskip 0.5cm
\begin{center}
\epsfig{figure=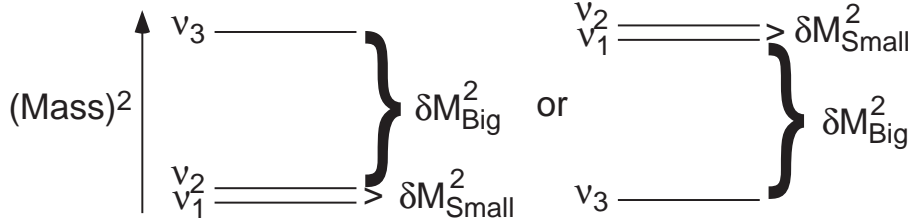,width=12cm} 
\end{center}
\caption{A three neutrino (Mass)$^2$ spectrum in which the $\nu_2 - \nu_1$ splitting $\delta M^2_{\mbox{Small}}$ is much smaller than the splitting $\delta M^2_{\rm Big}$ between $\nu_3$ and the $\nu_2 - \nu_1$ pair. The latter pair may be at either the bottom or the top of the spectrum.
\label{f4}}
\end{figure}

Suppose that an oscillation experiment has $L/E$ such that $\delta M^2_{\rm Big} L\,/E$ is of order unity, which implies that $\delta M^2_{\rm Small} L/E \ll 1$. For this experiment, and $\beta \neq \alpha$, the oscillation amplitude of \eqn{25} is given approximately by 
\be
e^{i M_3^2 \frac{L}{2E}} A(\nu_\alpha \ra \nu_{\beta\neq\alpha}) 
\cong (U^*_{\alpha 1} U_{\beta 1} + U^*_{\alpha 2} U_{\beta 2}) 
e^{i \delta M^2_{32} \frac{L}{2E}} + U^*_{\alpha 3} U_{\beta 3}\; .
\label{eq40}
\ee
Using the unitarity constraint of \eqn{27}, this becomes
\bea
e^{i M_3^2 \frac{L}{2E}} A(\nu_\alpha \ra \nu_{\beta\neq\alpha}) 
& \cong & U^*_{\alpha 3} U_{\beta 3} (1 -  e^{i \delta M^2_{32} \frac{L}{2E}})  \nonumber \\
& = & -2i e^{i \delta M^2_{32} \frac{L}{4E}} 
	U^*_{\alpha 3} U_{\beta 3} \sin(\delta M^2_{32}\frac{L}{4E})\; .
\label{eq41}
\eea
Taking the absolute square of this relation, using $|\delta M^2_{32}| \equiv \delta M^2_{\rm Big}$, and inserting the omitted factors of $\hbar$ and $c$, we find that the $\nu_\alpha \ra \nu_{\beta\neq\alpha}$ oscillation probability is given by
\be
P(\nu_\alpha \ra \nu_{\beta\neq\alpha}) =  
	4|U_{\alpha 3}|^2 |U_{\beta 3}|^2 \sin^2 (1.27 \,\delta M^2_{\rm Big} (\mbox{eV}^2) \frac{L\mbox{(km)}}{E\mbox{(GeV)}}) \; .
\label{eq42}
\ee
To find the corresponding probability $P(\nu_\alpha \ra \nu_\alpha)$ that a neutrino of flavor $\alpha$ retains its original flavor, we simply use the conservation of probability, \eqn{30}:
\be
	P(\nu_\alpha \ra \nu_\alpha) = 1 - \sum_{\beta\neq\alpha} 
	P(\nu_\alpha \ra \nu_\beta) \; .
\label{eq43}
\ee
From \eqn{42} and the unitarity relation $\sum_{\beta\neq\alpha} |U_{\beta 3}|^2 = 1 - |U_{\alpha 3}|^2$, we then find that
\be
P(\nu_\alpha \ra \nu_\alpha) = 1 - 4|U_{\alpha 3}|^2 (1-|U_{\alpha 3}|^2) \sin^2 (1.27 \,\delta M^2_{\rm Big} (\mbox{eV}^2) \frac{L\mbox{(km)}}{E\mbox{(GeV)}}) \; .
\label{eq44}
\ee

Comparing Eqs.~(\ref{eq42}) and (\ref{eq34}), and Eqs.~(\ref{eq44}) and (\ref{eq36}), we see that in the three-neutrino scenario with $\delta M^2_{\rm Small} L/E \ll 1$, the oscillation probabilities are the same as in the two-neutrino case, except that with three neutrinos, the isolated one, $\nu_3$, plays the role played by $\nu_2$ when there are only two neutrinos. 
This strong similarity between the oscillation probabilities in the two cases is easy to understand. When the three-neutrino case is studied by an experiment for which $\delta M^2_{\rm Small} L/E \ll 1$, the experiment cannot see the splitting between $\nu_1$ and $\nu_2$. In such an experiment, there appear to be only two neutrinos with distinct masses: the $\nu_1$ and $\nu_2$ pair, which looks like a single neutrino, and $\nu_3$. As a result, the two neutrino oscillation probabilities hold.\cite{b6}

\section{Neutrino Oscillation in Matter}

So far, we have been talking about neutrino oscillation {\em in vacuum}. However, some very important oscillation experiments are concerned with neutrinos that travel through a lot of matter before reaching the detector. These neutrinos include those made by nuclear reactions in the core of the sun, which traverse a lot of solar material on their way out of the sun towards solar neutrino detectors here on earth. 
They also include the neutrinos made in the earth's atmosphere by cosmic rays. These atmospheric neutrinos can be produced in the atmosphere on one side of the earth, and then travel through the whole earth before being detected in a detector on the other side. To deal with the solar and atmospheric neutrinos, we need to understand how passage through matter affects neutrino oscillation.

To be sure, the interaction between neutrinos and matter is extremely feeble. Nevertheless, the coherent forward scattering of neutrinos from many particles in a material medium can build up a big effect on the oscillation amplitude.

For both the solar and atmospheric neutrinos, it is a good approximation to take just two neutrinos into account. Furthermore, it is convenient to treat the propagation of neutrinos in matter in terms of an effective Hamiltonian. To set the stage for this treatment, let us first derive the Hamiltonian for travel through vacuum. For the sake of illustration, let us suppose that the two neutrino flavors that need to be considered are $\nu_e$ and $\nu_\mu$. The most general time-dependent neutrino state vector $\ket{\nu(t)}$ can then be written as
\be
\ket{\nu(t)} = \sum_{\alpha=e,\mu} f_\alpha (t) \ket{\nu_\alpha} \; ,
\label{eq2.1}
\ee
where $f_\alpha (t)$ is the time-dependent amplitude for the neutrino to have flavor $\alpha$. If ${\cal H}_V$ (short for ${\cal H}_{\rm Vacuum}$) is the Hamiltonian for this two-neutrino system in vacuum, then Schr\"{o}dinger's equation for $\ket{\nu(t)}$ reads
\bea
i\frac{\partial}{\partial t}\ket{\nu(t)} & = & \sum_\alpha i \dot{f}_\alpha (t) \ket{\nu_\alpha} \nonumber \\
&=& {\cal H}_V \ket{\nu(t)} = \sum_\beta f_\beta (t) {\cal H}_V \ket{\nu_\beta} \nonumber \\
&=& \sum_\beta f_\beta (t) \sum_\alpha\ket{\nu_\alpha} \bra{\nu_\alpha}{\cal H}_V \ket{\nu_\beta} \nonumber \\
&=& \sum_\alpha\left[ \sum_\beta({\cal H}_V)_{\alpha\beta}f_\beta(t) \right] \ket{\nu_\alpha} \; .
\label{eq2.2}
\eea
Here, $({\cal H}_V)_{\alpha\beta} \equiv \bra{\nu_\alpha}{\cal H}_V \ket{\nu_\beta}$. Comparing the coefficients of $\ket{\nu_\alpha}$ at the beginning and end of \eqn{2.2}, we clearly have
\be
i\frac{\partial}{\partial t} \left[ \begin{array}{c}
	f_e (t) \\ f_\mu (t) \end{array} \right] = {\cal H}_V
	\left[ \begin{array}{c} f_e (t) \\ f_\mu (t) \end{array} \right] 
	\;\; ,
\label{eq2.3}
\ee
where ${\cal H}_V$ is now the $2\times 2$ matrix with elements $({\cal H}_V)_{\alpha\beta}$. The Schr\"{o}dinger equation (\ref{eq2.3}) is completely analogous to the familiar one for a spin-$1/2$ particle. The roles of the two spin states are now being played by the two flavor states.

Let us call the two neutrino mass eigenstates out of which $\nu_e$ and $\nu_\mu$ are made $\nu_1$ and $\nu_2$. To find the matrix ${\cal H}_V$, let us assume that our neutrino has a definite momentum $p$, so that its mass-eigenstate component $\nu_i$ has definite energy $E_i$ given by \eqn{21}. That is, ${\cal H}_V \ket{\nu_i} = E_i \ket{\nu_i}$, and the different mass eigenstates $\ket{\nu_i}$, like the eigenstates of any Hermitean Hamiltonian, are orthogonal to each other. Then, in view of \eqn{11}, the elements $({\cal H}_V)_{\alpha\beta}$ of the vacuum Hamiltonian are given by
\be
({\cal H}_V)_{\alpha\beta} \equiv \bra{\nu_\alpha}{\cal H}_V\ket{\nu_\beta} = \bra{\sum_i U^*_{\alpha i}\nu_i}{\cal H}_V\ket{\sum_j U^*_{\beta j}\nu_j} \nonumber \\
= \sum_i U_{\alpha i} U^*_{\beta i} E_i \; .
\label{eq2.4}
\ee
In this expression, the two-neutrino mixing matrix $U$ may be taken from \eqn{37}. However, we have seen that the complex phase factors in \eqn{37} have no effect on the two-neutrino oscillation probabilities, \eqn{38}. Indeed, it is not hard to show that when there are only two neutrinos, complex phases in $U$ have no effect whatsoever on neutrino oscillation. Thus, since oscillation is our only concern here, we may remove the complex phase factors from the $U$ of \eqn{37}. If, in addition, we relabel the mixing angle $\theta_V$ (short for $\theta_{\rm Vacuum}$), $U$ becomes
\be
U = \left[ \begin{array}{c@{\hspace{.15in}}c}
		\cos\theta_V  &  \sin\theta_V \\ 
		-\sin\theta_V &  \cos\theta_V  \end{array} \right] \;\; .
\label{eq2.5}
\ee
Inserting in \eqn{2.4} the elements $U_{\alpha i}$ of this matrix and the energies $E_i$ given by \eqn{21}, we can obtain all the $({\cal H}_V)_{\alpha\beta}$.

The matrix ${\cal H}_V$ can be put into a more symmetric and convenient form if we add to it a suitably chosen multiple $(\Delta E)I$ of the identity matrix $I$. Such an addition will not change the predictions of ${\cal H}_V$ for neutrino oscillation. To see why, we note first that the identity matrix is invariant under the unitary transformation that diagonalizes ${\cal H}_V$. 
Thus, adding $(\Delta E)I$ to ${\cal H}_V$ in the flavor basis, where its elements are $({\cal H}_V)_{\alpha\beta}$, is equivalent to adding $(\Delta E)I$ to ${\cal H}_V$ in the mass eigenstate basis, where it is diagonal. Hence, if the eigenvalues of ${\cal H}_V$ are $E_i, i=1,2$, those of ${\cal H}_V + (\Delta E)I$ are $E_i + \Delta E, i=1,2$. That is, both eigenvalues are displaced by the same amount, $\Delta E$. To see that such a common shift of all eigenvalues does not affect neutrino oscillation, suppose a neutrino is born at time $t=0$ with flavor $\alpha$. That is, $\ket{\nu(0)} = \ket{\nu_\alpha}$. After a time $t$, this neutrino will have evolved into the state $\ket{\nu(t)}$ given, according to Schr\"{o}dinger's equation, by
\bea
\ket{\nu(t)} & = & e^{-i{\cal H}_V t} \ket{\nu(0)} \nonumber \\
&=& e^{-i{\cal H}_V t} \sum_i U^*_{\alpha i}\ket{\nu_i}
=  \sum_i U^*_{\alpha i} e^{-iE_i t}\ket{\nu_i} \;\; .
\label{eq2.6}
\eea
The amplitude $A(\nu_\alpha \ra \nu_\beta)$ for this neutrino to have oscillated into a $\nu_\beta$ in the time $t$ is then given by
\be
A(\nu_\alpha \ra \nu_\beta) = \bra{\nu_\beta}\nu(t)\rangle
= \sum_i U^*_{\alpha i} e^{-iE_i t} U_{\beta i} \;\; .
\label{eq2.7}
\ee
Clearly, if we add $(\Delta E)I$ to ${\cal H}_V$ so that its eigenvalues $E_i$ are replaced by $E_i + \Delta E$, then $A(\nu_\alpha \ra \nu_\beta)$ is just multiplied by the overall phase factor $\exp [-i(\Delta E)t]$. Obviously, this phase factor has no effect on the oscillation probability $P(\nu_\alpha \ra \nu_\beta)  =  |A(\nu_\alpha \ra \nu_\beta)|^2$. Thus, the addition of $(\Delta E)I$ to ${\cal H}_V$ does not affect neutrino oscillation.

For our purposes, the most convenient choice of $\Delta E$ is $-[p+(M^2_1 + M^2_2)/4p]$. Then, from \eqn{2.4}, the new effective Hamiltonian ${\cal H}_V^\prime \equiv {\cal H}_V + (\Delta E)I$ has the matrix elements
\be
({\cal H}_V^\prime)_{\alpha\beta} = \sum_i U_{\alpha i}U^*_{\beta i} E_i - [p + \frac{M^2_1 + M^2_2}{4p}] \delta_{\alpha\beta} \;\; .
\label{eq2.8}
\ee
From Eqs.~(\ref{eq2.5}) and (\ref{eq21}), this gives\cite{b2.1}
\be
{\cal H}_V^\prime = \frac{\dm21}{4E}
	\left[ \begin{array}{c@{\hspace{.15in}}c}
	-\cos2\theta_V  &  \sin2\theta_V \\ 
	\sin 2\theta_V &  \cos 2\theta_V  \end{array} \right] \;\; .
\label{eq2.9}
\ee
Here, we have used the fact that $p \cong E$, the energy of the neutrino averaged over its two mass-eigenstate components. We leave to the reader the instructive exercise of verifying that, inserted into the Schr\"{o}dinger \eqn{2.3}, the ${\cal H}_V^\prime$ of \eqn{2.9} does indeed lead to the usual two-neutrino oscillation probability, \eqn{38}.

With the Hamiltonian that governs neutrino propagation through the vacuum in hand, let us now ask how neutrino propagation is modified by the presence of matter. Matter, of course, consists of electrons and nucleons. When passing through a sea of electrons and nucleons, a (non-sterile) neutrino can undergo the forward elastic scatterings depicted in Fig.~(\ref{f5}).
\begin{figure}[t]
\vskip 0.5cm
\begin{center}
\epsfig{figure=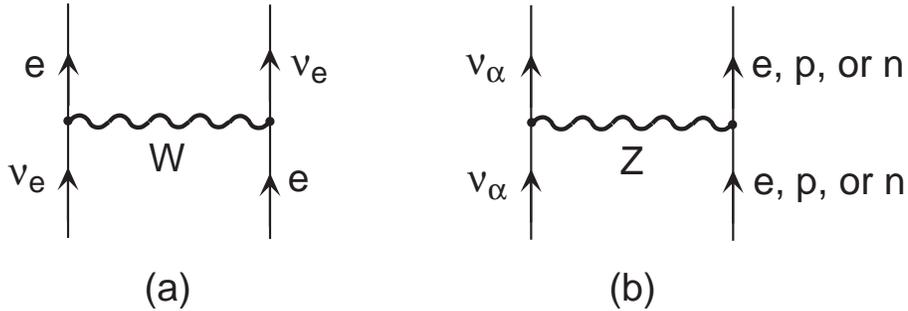,width=12cm} 
\end{center}
\caption{Forward elastic scattering of a neutrino from a particle of matter. (a) W-exchange-induced scattering from an electron, which is possible only for a $\nu_e$. (b) Z-exchange-induced scattering from an electron, proton, or neutron. This is possible for $\nu_\alpha = \nu_e, \nu_\mu$, or $\nu_\tau$. According to the Standard Model, the amplitude for this Z exchange is the same, for any given target particle, for all three active neutrino flavors.
\label{f5}}
\end{figure}
Coherent forward scatterings, via the pictured processes, from many particles in a material medium will give rise to an interaction potential energy of the neutrino in the medium. Since one of the reactions in Fig.~(\ref{f5}) can occur only for electron neutrinos, this interaction potential energy will depend on whether the neutrino is a $\nu_e$ or not. The interaction potential energy for a neutrino of flavor $\alpha$ must be added to the matrix element $({\cal H}_V^\prime)_{\alpha\alpha}$ to obtain the Hamiltonian for propagation of a neutrino in matter.\cite{b2.2}

An important application of this physics is to the motion of solar neutrinos through solar material. The solar neutrinos are produced in the center of the sun by nuclear reactions such as $p+p \ra$ $^2\!H + e^+ + \nu_e$. The neutrinos produced by these reactions are all electron neutrinos. Let us suppose that the only neutrinos with which electron neutrinos mix appreciably are muon neutrinos, so that we have a two-neutrino system of the kind we have just been discussing. 
The solar neutrinos stream outward from the center of the sun in all directions, some of them eventually arriving at solar neutrino detectors here on earth. The passage of these solar neutrinos through solar material on their way out of the sun modifies their oscillation. Any neutrino which is still a $\nu_e$, as it was at birth, can interact with solar electrons via the W exchange of Fig.~(5a). This interaction leads to an interaction potential energy $V_W(\nu_e)$ of an electron neutrino in the sun. 
This $V_W(\nu_e)$ is obviously proportional to the Fermi coupling constant $G_F$, which governs the amplitude for the process in Fig.~(5a). It is also proportional to the number of electrons per unit volume, $N_e$, at the location of the neutrino, since $N_e$ measures the number of electrons which can contribute coherently to the forward $\nu_e$ scattering. One can show\cite{b2.3} that in the Standard Model,
\be
V_W(\nu_e) = \sqrt{2} G_F N_e \;\; .
\label{eq2.10}
\ee
This energy must be added to $({\cal H}_V^\prime)_{ee}$ to obtain the Hamiltonian for propagation of a neutrino in the sun.

In principle, the interaction energy produced by the Z exchanges of Fig.~(5b) must also be added to ${\cal H}_V^\prime$. However, since these Z exchanges are both flavor diagonal and flavor independent, their contribution to the Hamiltonian is a multiple of the identity matrix. As we have already seen, a contribution of this character does not affect neutrino oscillation. Thus, we may safely ignore it.

In incorporating $V_W(\nu_e)$ into the Hamiltonian, it is convenient, in the interest of symmetry, to add as well the multiple $-\frac{1}{2} V_W(\nu_e) I$ of the identity matrix. Thus, with ${\cal H}_V^\prime$ the Hamiltonian for propagation in vacuum given by \eqn{2.9}, the Hamiltonian ${\cal H}_\odot$ for propagation in the sun is given by\cite{b2.2}
\bea
{\cal H}_\odot & = & {\cal H}_V^\prime + \frac{G_F}{\sqrt{2}} N_e
	\left[ \begin{array}{c@{\hspace{.15in}}c}
		1 & 0 \\ 0 & -1 \end{array} \right] \nonumber \\
	& = & \frac{\dm21 D_\odot}{4E} 
	\left[ \begin{array}{c@{\hspace{.15in}}c}
	-\cos2\theta_\odot  &  \sin2\theta_\odot \\ 
	\sin 2\theta_\odot &  \cos 2\theta_\odot  \end{array} \right] \;\; .
\label{eq2.11}
\eea
In this expression,
\be
D_\odot = \sqrt{\sin^2 2\theta_\odot + (\cos 2\theta_\odot - x_\odot)^2} \; ,
\label{eq2.12}
\ee
and
\be
\sin^2 2\theta_\odot = \frac{\sin^2 2\theta_V}{\sin^2 2\theta_V + (\cos 2\theta_V - x_\odot)^2} \; ,
\label{eq2.13}
\ee
where
\be
x_\odot = \frac{2\sqrt{2} G_F N_e E}{\dm21} \;\; .
\label{eq2.14}
\ee
The angle $\theta_\odot$ is the effective neutrino mixing angle in the sun when the electron density is $N_e$.

We note that ${\cal H}_\odot$, \eqn{2.11}, has precisely the same form as ${\cal H}_V^\prime$, \eqn{2.9}. The only difference between these two Hamiltonians is that the parameters---the mixing angle and the effective neutrino (Mass)$^2$ splitting out in front of the matrix---have different values. Of course, the electron density $N_e$ is not a constant, but depends on the distance $r$ from the center of the sun. 
Thus, the parameters $\theta_\odot$ and $(\dm21)D_\odot$ are not constant either, unlike their counterparts, $\theta_V$ and $\dm21$, in ${\cal H}_V^\prime$. However, let us imagine for a moment that $N_e$ {\em is} a constant. Then $H_\odot$, like ${\cal H}_V^\prime$, is independent of position, and must lead to the same oscillation probability, \eqn{38}, as ${\cal H}_V^\prime$ does, except for the substitutions $\theta \ra \theta_\odot$ and $\dm21 \ra (\dm21)D_\odot$. That is, in matter of constant electron density $N_e$, $H_\odot$ leads to the oscillation probability
\bea
P(\nu_e \ra \nu_\mu) & = & P(\nu_\mu \ra \nu_e) \nonumber \\
	& = & \sin^2 2\theta_\odot \sin^2 [1.27 \,\delta M^2_{21} (\mbox{eV}^2) D_\odot \frac{L\mbox{(km)}}{E\mbox{(GeV)}}] \;\; .
\label{eq2.15}
\eea

Now let $N_e$ vary with $r$ as it does in the real world. However, suppose that it varies slowly enough that the constant-$N_e$ picture we have just painted applies at any given radius $r$, but with $N_e(r)$, hence $x_\odot(r)$, slowly decreasing as $r$ increases. 
Suppose also that $\dm21$ and $E$ are such that $x_\odot(r=0) > \cos2\theta_V$. Then, assuming that $\cos 2\theta_V > 0$, there must be a radius $r = r_c$ somewhere between $r=0$ and the outer edge of the sun, where $N_e \ra 0$, such that $x_\odot(r_c) = \cos2\theta_V$. From \eqn{2.13}, we see that at this special radius $r_c$, there is a kind of ``resonance'' with $\sin^2 2\theta_\odot = 1$, even if $\theta_V$ is tiny. That is, mixing can be maximal in the sun even if it is very small in vacuo. As a result, the oscillation probability, which is proportional to the mixing factor $\sin^2 2\theta_\odot$ as we see in \eqn{2.15}, can be very large.

A nice picture of this enhanced probability for flavor transitions in matter can be gained by considering the neutrino energy eigenvalues and eigenvectors. If we neglect the inconsequential Z exchange contribution, and take $H_\odot$ from \eqn{2.11}, then the true Hamiltonian for propagation in the sun is
\be
{\cal H}_{\rm True} = {\cal H}_\odot + [p + \frac{M^2_1 + M^2_2}{4p} + \frac{1}{2}V_W(\nu_e)] \, I \;\; ,
\label{eq2.16}
\ee
since the second term in this expression was subtracted from the true Hamiltonian to get ${\cal H}_\odot$. Now, $p+(M^2_1 +M^2_2)/4p \cong E$, the energy our neutrino would have in vacuum, averaged over its two mass-eigenstate components. Thus, from Eqs.~(\ref{eq2.16}) and (\ref{eq2.11}), (\ref{eq2.9}), (\ref{eq2.10}), and (\ref{eq2.14}), we have in the $\nu_e-\nu_\mu$ basis 
\be
{\cal H}_{\rm True}(r) = \left[ \begin{array}{cc}
	E & 0 \\ 0 & E \end{array} \right] + \frac{\dm21}{4E} 
	\left[ \begin{array}{c@{\hspace{.15in}}c}
	-\cos2\theta_V+2x_\odot(r)  &  \sin2\theta_V \\ 
	\sin 2\theta_V   &  \cos 2\theta_V  \end{array} \right] \;\; .
\label{eq2.17}
\ee
If we continue to assume that $N_e(r)$, and consequently $x_\odot(r)$, varies slowly, we may diagonalize this Hamiltonian for one $r$ at a time to see how a solar neutrino will behave. We find  from \eqn{2.17} that for a given $r$, the energy eigenvalues $E_\pm(r)$ are given by
\be
E_\pm(r) = E + \frac{\dm21}{4E}\left[ x_\odot(r) \pm \sqrt{(x_\odot(r)-\cos2\theta_V)^2 + \sin^2 2\theta_V}\right]\;\; .
\label{eq2.18}
\ee
To explore the implications of these energy levels, let us suppose that $\dm21$ is such that at $r=0$, where $N_e(r)$ and hence $x_\odot(r)$ has its maximum value, $x_\odot \gsim 1$. From \eqn{2.14}, the value of $G_F$, the value ($\sim 10^{26}$/cc) of $N_e(r=0)$, and the typical energy $E$ ($\sim 1$ MeV) of a solar neutrino, we find that the required $\dm21$ is of order $10^{-5}$ eV$^2$. 
The dominant term in the energies $E_\pm(r)$ of \eqn{2.18} will be the first one, $E$ ($\sim 1$ MeV). However, very interesting physics will result from the second term, despite the fact that this term is only of order $\dm21/4E \sim (10^{-5}{\rm eV}^2) / 1 {\rm MeV} \sim 10^{-17}\; $ MeV!

The neutrino states which propagate in the sun without mixing significantly with each other are the eigenvectors of ${\cal H}_{\rm True}(r)$. To study these eigenvectors, let us assume for simplicity that the vacuum mixing angle $\theta_V$ is small. Then it quickly follows that, except in the vicinity of the special radius $r_c$ where $x_\odot(r_c) = \cos2\theta_V$, one of the eigenvectors of ${\cal H}_{\rm True}(r)$, \eqn{2.17}, is essentially pure $\nu_e$, while the other is essentially pure $\nu_\mu$. The evolution of a neutrino traveling outward through the sun is then as depicted in Fig.~(\ref{f6}). The neutrino follows the trajectory indicated by the arrows. 
\begin{figure}[t]
\vskip 0.5cm
\begin{center}
\epsfig{figure=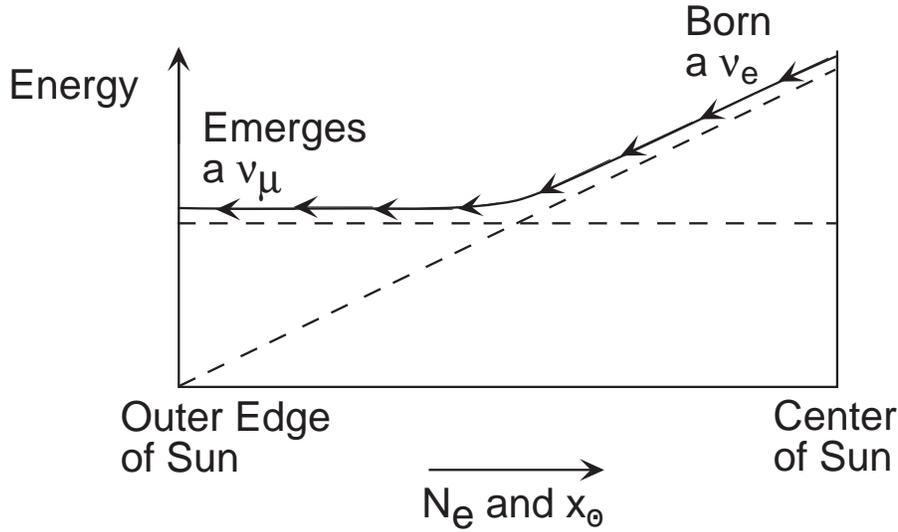,width=12cm} 
\end{center}
\caption{Propagation of a neutrino in the sun. The horizontal axis is linear in $N_e$ and $x_\odot$. The solid line shows the upper eigenvalue, $E_+(r)$, for small $\theta_V$. The dashed lines show the two eigenvalues, $E_\pm(r)$, when there is no vacuum mixing ($\theta_V = 0)$. A neutrino born as a $\nu_e$ follows the path indicated by the arrows.
\label{f6}}
\end{figure}
Produced by some nuclear process, the neutrino is born at small $r$ as a $\nu_e$. From \eqn{2.17}, the eigenvector that is essentially $\nu_e$ at $r=0$ is the one with the higher energy, $E_+(r)$. Thus, our neutrino begins its outward journey through the sun as the eigenvector belonging to the upper eigenvalue, $E_+$. Since the eigenvectors do not cross at any $r$ and do not mix appreciably, the neutrino will remain this eigenvector. 
However, the flavor content of this eigenvector changes dramatically as the neutrino passes through the region near the radius $r=r_c$ where $x_\odot(r) - \cos2\theta_V = 0$. For $r \ll r_c$, the eigenvector belonging to $E_+(r)$ is essentially a $\nu_e$, as one may see from \eqn{2.17} if one neglects the small off-diagonal terms proportional to $\sin 2\theta_V$, and imposes the small-$r$ condition $x_\odot > \cos2\theta_V$. But for $r \gg r_c$, the eigenvector belonging to the higher energy level $E_+(r)$ is essentially a $\nu_\mu$, as one may also see from \eqn{2.17} if one neglects the off-diagonal terms and imposes the large-$r$ condition $x_\odot < \cos2\theta_V$. 
Thus, the eigenvector corresponding to the higher energy eigenvalue $E_+(r)$, along which the solar neutrino travels, starts out as a $\nu_e$ at the center of the sun, but ends up as a $\nu_\mu$ at the outer edge of the sun. The solar neutrino, born a $\nu_e$ in the solar core, emerges from the rim of the sun as a $\nu_\mu$. Furthermore, it does this with high probability even if the vacuum mixing angle $\theta_V$ is very small, so that oscillation in vacuum would not have much of an effect. This very efficient conversion of solar electron neutrinos into neutrinos of another flavor as a result of interaction with matter is known as the Mikheyev-Smirnov-Wolfenstein (MSW) effect.\cite{b2.2,b2.4}

We turn now from solar neutrinos to atmospheric neutrinos, whose propgation through the earth entails a second important application of the physics of neutrinos traveling through matter. As already mentioned, an atmospheric neutrino can be produced in the atmosphere on one side of the earth, and then journey through the whole earth to be detected in a detector on the other side.  While traveling through the earth, this neutrino will undergo interactions that can significantly modify its oscillation pattern.

There is very strong evidence\cite{b2.5} that the atmospheric neutrinos born as muon neutrinos oscillate into neutrinos $\nu_x$ of another flavor. It is known that $\nu_x$ is not a $\nu_e$. It could be a $\nu_\tau$, or a sterile neutrino $\nu_s$, or sometimes one of these and sometimes the other. One way to find out what is becoming of the oscillating muon neutrinos is to see whether their oscillation is afffected by their passage through earth-matter. The oscillation $\nu_\mu \ra \nu_s$ will be affected, but $\nu_\mu \ra \nu_\tau$ will not be. 

To see why this is so, let us first assume that the oscillation is $\nu_\mu \ra \nu_\tau$. Then the Hamiltonian ${\cal H}_E$ (short for ${\cal H}_{\rm Earth}$) that describes neutrino propagation in the earth is a $2\times 2$ matrix in $\nu_\mu - \nu_\tau$ space. Now, either a $\nu_\mu$ or a $\nu_\tau$ will interact with earth-matter via the Z exchange of Fig.~(5b). 
This interaction will give rise to an interaction potential energy of the neutrino in the earth. However, according to the S(tandard) M(odel), the Z-exchange amplitude is the same for a $\nu_\tau$ as it is for a $\nu_\mu$. Thus, in the case of $\nu_\mu \ra \nu_\tau$ oscillation, the contribution of neutrino-matter interaction to ${\cal H}_E$ is a multiple of the identity matrix. As we have alrady seen, such a contribution has no effect on oscillation.

Now, suppose the oscillation is not $\nu_\mu \ra \nu_\tau$ but $\nu_\mu \ra \nu_s$. Then ${\cal H}_E$ is a $2\times 2$ matrix in $\nu_\mu - \nu_s$ space. A $\nu_\mu$ will interact with earth-matter, as we have discussed, but a $\nu_s$, of course, will not. Thus, the contribution of neutrino-matter interaction to ${\cal H}_E$ is now of the form
\be
  \begin{array}{c@{\hspace{.15in}}c}
& \begin{array}{c@{\hspace{.15in}}c}   
\nu_\mu & \nu_s \end{array} \\ [.2pc]
\begin{array}{cc} \nu_\mu \\ \nu_s \end{array}
  & \left[ \begin{array}{c@{\hspace{.15in}}c}
	V_Z(\nu_\mu) & 0 \\ 0 & 0 \end{array}\right] \;\; ,
	\end{array}
\label{eq2.19}
\ee
where $V_Z(\nu_\mu)$ is the interaction potential energy of muon neutrinos produced by the Z exchange of Fig.~(5b). Since the matrix (\ref{eq2.19}) is not a multiple of the identity, neutrino-matter interaction does affect $\nu_\mu \ra \nu_s$ oscillation.

According to the SM, the forward Z-exchange amplitudes for a target $e$ and a target $p$ are equal and opposite. Thus, assuming that the earth is electrically neutral so that it contains an equal number of electrons and protons per unit volume, the $e$ and $p$ contributions to $V_Z(\nu_\mu)$ cancel. Then $V_Z(\nu_\mu)$ is proportional to the neutron number density, $N_n$. Taking the proportionality constant from the SM, we have
\be
V_Z(\nu_\mu) = - \frac{G_F}{\sqrt{2}} N_n  \;\; .
\label{eq2.20}
\ee
To obtain the Hamiltonian ${\cal H}_E$ for $\nu_\mu \ra \nu_s$ oscillation in the earth, we add to the vacuum Hamiltonian ${\cal H}_V^\prime$ of \eqn{2.9} the contribution (\ref{eq2.19}) from matter interactions, using \eqn{2.20} for $V_Z(\nu_\mu)$. 
{\em Of course, it must be understood that ${\cal H}_V^\prime$ is now to be taken as a matrix in $\nu_\mu - \nu_s$ space, and that the vacuum (Mass)$^2$ splitting $\dm21$ and mixing angle $\theta_V$ in ${\cal H}_V^\prime$ are now different parameters than they were when we obtained from ${\cal H}_V^\prime$ the Hamiltonian ${\cal H}_\odot$ for neutrino propagation in the sun. The quantities $\dm21$ and $\theta_V$ are now new parameters appropriate to the vacuum oscillation of atmospheric, rather than solar, neutrinos.} 
To obtain a more symmetrical and convenient ${\cal H}_E$ from ${\cal H}_V^\prime$, we also add $-\frac{1}{2}V_Z(\nu_\mu)I$, a multiple of the identity which will not affect the implications of ${\cal H}_E$ for oscillation. The result is 
\be
{\cal H}_E = \frac{(\dm21)D_E}{4E} 
	\left[ \begin{array}{c@{\hspace{.15in}}c}
	-\cos2\theta_E  &  \sin2\theta_E \\ 
	\sin 2\theta_E  &  \cos 2\theta_E  \end{array} \right] \;\; .
\label{eq2.21}
\ee
Here, 
\be
D_E =  \sqrt{\sin^2 2\theta_V + (\cos 2\theta_V - x_E)^2} \;\; ,
\label{eq2.22}
\ee
and
\be
\sin^2 2\theta_E = \frac{\sin^2 2\theta_V}{\sin^2 2\theta_V + (\cos 2\theta_V - x_E)^2} \; ,
\label{eq2.23}
\ee
where
\be
x_E = - \frac{\sqrt{2} G_F N_n E}{\dm21} \;\; .
\label{eq2.24}
\ee

As a rough approximation, we may take the neutron density $N_n$ to be constant throughout the earth. Then, $x_E$ is also a constant, and the Hamiltonian ${\cal H}_E$ of \eqn{2.21} is identical to the vacuum Hamiltonian ${\cal H}_V^\prime$ of \eqn{2.9}, except that the constant $\dm21$ is replaced by the constant $(\dm21)D_E$, and the constant $\theta_V$ by the constant $\theta_E$. Thus, from the fact that ${\cal H}_V^\prime$ leads to the vacuum oscillation probability of \eqn{38} (with $\nu_e \ra \nu_\mu$ replaced by $\nu_\mu \ra \nu_s$ for the present application), we immediately conclude that the ${\cal H}_E$ of \eqn{2.21} leads to the oscillation probability 
\be
P(\nu_\mu \ra \nu_s) = \sin^2 2\theta_E \sin^2 [1.27 \,\dm21 (\mbox{eV}^2) D_E \frac{L\mbox{(km)}}{E\mbox{(GeV)}}] \;\; .
\label{eq2.25}
\ee

As we see from \eqn{2.24}, $x_E$, which is a measure of the influence of matter effects on atmospheric neutrinos, grows with energy $E$. In a moment we will see that for $E \sim 1$ GeV, matter effects are negligible. Fits to data\cite{b2.5a} on atmospheric neutrinos with roughly this energy have led to the conclusion that atmospheric neutrino oscillation involves a neutrino (Mass)$^2$ splitting $\delta M^2_{\rm Atmos}$ given by
\be
	\delta M^2_{\rm Atmos} \sim 3\times 10^{-3} \mbox{eV}^2 \;\; ,
\label{eq2.26}
\ee
and a neutrino mixing angle $\theta_{\rm Atmos}$ given by
\be
	\sin^2 2\theta_{\rm Atmos} \sim 1 \;\; .
\label{eq2.27}
\ee
That is, the mixing when matter effects are negligible is very large, and perhaps maximal. The quantities $\delta M^2_{\rm Atmos}$ and $\theta_{\rm Atmos}$ are to be taken, respectively, for $\dm21$ and $\theta_V$ in Eqs.~(\ref{eq2.21}) - (\ref{eq2.25}) to find the implications of those equations for $\nu_\mu \ra \nu_s$ within the earth.

Since atmospheric neutrino oscillation involves maximal mixing when matter effects are negligible, the matter effects cannot possibly enhance the oscillation, but can only suppress it. From \eqn{2.23}, we see that if, as observed, $\sin^2 2\theta_V \simeq 1$, then matter effects will lead to a smaller effective mixing $\sin^2 2\theta_E$ in the earth, given by 
\be
	\sin^2 2\theta_E = \frac{1}{1 + x^2_E} \;\; .
\label{eq2.28}
\ee
As is clear in \eqn{2.25}, this will result in a smaller oscillation probability $P(\nu_\mu \ra \nu_s)$ than one would have in vacuum, where $\sin^2 2\theta_E$ is replaced by $\sin^2 2\theta_V$ ($\sim1$). 

Since $x_E$ grows with energy, the degree to which matter effects suppress $\nu_\mu \ra \nu_s$ grows as well. From \eqn{2.24} for $x_E$, the known values of $G_F$ and $N_n$, and the value (\ref{eq2.26}) required for $\dm21$ by the data, we find that $x_E \ll 1$ when $E \sim 1$ GeV. Thus, matter effects are indeed negligible at this energy, so it is legitimate to determine\cite{b2.5a} the vacuum parameters $\delta M^2_{\rm Atmos}$ and $\sin^2 2\theta_{\rm Atmos}$ by analyzing the $\sim 1$ GeV data neglecting matter effects. However, at sufficiently large $E$, the matter-induced suppression of $\nu_\mu \ra \nu_s$ will obviously be significant. From \eqn{2.24}, we find that $\sin^2 2\theta_E$ is below 1/2 when $E \gsim 20$ GeV. The consequent suppression of oscillation at these energies has been looked for, and is not seen.\cite{b2.6} This absence of suppression is a powerful part of the evidence that the neutrinos into which the atmospheric muon neutrinos oscillate are not sterile neutrinos, or at least not solely  sterile neutrinos.\cite{b2.6}

\section{Conclusion}

Evidence has been reported that the solar neutrinos, the atmospheric neutrinos, the accelerator-generated neutrinos studied by the Liquid Scintillator Neutrino Detector (LSND) experiment at Los Alamos, and the accelerator-generated neutrinos studied by the K2K experiment in Japan, actually do oscillate. Some of this evidence is very strong. The neutrino oscillation experiments, present and future, are discussed in this Volume by John Wilkerson.\cite{b2.7}

In these lectures, we have tried to explain the basic physics that underlies neutrino oscillation, and that is invoked to understand the oscillation experiments. As we have seen, neutrino oscillation implies neutrino mass and mixing. Thus, given the compelling evidence that at least some neutrinos do oscillate, we now know that neutrinos almost certainly have nonzero masses and mix. This knowledge rasises a number of questions about the neutrinos:

\begin{itemize}
	\item How many neutrino flavors, including both interacting and possible sterile flavors, are there? Equivalently, how many neutrino mass eigenstates are there?
	\item What are the masses, $M_i$, of the mass eigenstates $\nu_i$?
	\item Is the antiparticle $\overline{\nu_i}$ of a given mass eigenstate $\nu_i$ the same particle as $\nu_i$, or a different particle?
	\item What are the sizes and phases of the elements $U_{\alpha i}$ of the leptonic mixing matrix? Equivalently, what are the mixing angles and complex phase factors in terms of which $U$ may be described? Do complex phase factors in $U$ lead to CP violation in neutrino behavior?
	\item What are the electromagnetic properties of neutrinos? In particular, what are their dipole moments?
	\item What are the lifetimes of the neutrinos? Into what do they decay?
	\item What is the physics that gives rise to the masses, the mixings, and the other properties of the neutrinos?
\end{itemize}

Seeking the answers to these and other questions about the neutrinos will be an exciting adventure for years to come.

\section*{Acknowledgments} 
It is a pleasure to thank the organizers of TASI 2000 for an excellent summer school, and for giving me the opportunity to participate in it. I am grateful to Leo Stodolsky for a fruitful collaboration on the oscillations of both neutral mesons and neutrinos, and to Serguey Petcov and Lincoln Wolfenstein for a helpful discussion of neutrinos in matter. I am also grateful to my wife Susan for her accurate, patient, and gracious typing of the written version of these lectures.


\section*{References} 
\begin{thebibliography}{99}

\bibitem{b1} M. Gell-Mann, P. Ramond, and R. Slansky, in {\it Supergravity}, eds. D. Freedman and P. van Nieuwenhuizen (North Holland, Amsterdam, 1979) 315; T. Yanagida, in {\it Proceedings of the Workshop on Unified Theory and Baryon Number in the Universe,} eds. O. Sawada and A. Sugamoto (KEK, Tsukuba, Japan, 1979); R. Mohapatra and G. Senjanovi\'{c}, \Journal{\PRL}{44}{912}{1980}.

\bibitem{b2} C. Quigg, this Volume.

\bibitem{b3} Increasingly, $U$ is also being referred to as the ``Maki-Nakagawa-Sakata matrix'' in recognition of insightful early work reported in Z. Maki, M. Nakagawa, and S. Sakata, \Journal{\em Prog. Theor. Phys.}{28}{870}{1962}. 
For other pioneering work related to neutrino oscillation, see B. Pontecorvo, \Journal{\em Zh. Eksp. Teor. Fiz.}{53}{1717}{1967} [\Journal{\em Sov. Phys. JETP}{26}{984}{1968}]; V. Gribov and B. Pontecorvo, \Journal{\PLB}{28}{493}{1969}; S. Bilenky and B. Pontecorvo, \Journal{\em Phys. Reports C}{41}{225}{1978}; A. Mann and H. Primakoff, \Journal{\PRD}{15}{655}{1977}.

\bibitem{b4} Y. Srivastava, A. Widom, and E. Sassaroli, \Journal{\ZPC}{66}{601}{1995}.

\bibitem{b5} Y. Grossman and H. Lipkin, \Journal{\PRD}{55}{2760}{1997}; 
H. Lipkin, \Journal{\PLB}{348}{604}{1995}.

\bibitem{b6} B. Kayser and R. Mohapatra, to appear in {\it Current Aspects of Neutrino Physics}, ed. D. Caldwell (Springer-Verlag, Heidelberg, 2001).

\bibitem{b2.1} J. Bahcall, {\it Neutrino Astrophysics} (Cambridge Univ. Press, Cambridge, 1989). This book contains a nice discussion of neutrino oscillation in matter, and references to key original papers in the literature.

\bibitem{b2.2} The foundations of the physics and oscillations of neutrinos in matter are laid in L. Wolfenstein, \Journal{\PRD}{17}{2369}{1978}.

\bibitem{b2.3} See, for example, F. Boehm and P. Vogel, {\it Physics of Massive Neutrinos} (Cambridge Univ. Press, Cambridge, 1987).

\bibitem{b2.4} S. Mikheyev and A. Smirnov, \Journal{\em Sov. J. Nucl. Phys.}{42}{913}{1986}, \Journal{\em Sov. Phys. JETP}{64}{4}{1986}, \Journal{\em Nuovo Cimento}{9C}{17}{1986}.

\bibitem{b2.5} H. Sobel, \Journal{\NPB\ (Proc. Suppl.)}{91}{127}{2001}. See also W. Mann, {\it ibid.}, p.~134, and B. Barish, {\it ibid.}, p.~141.

\bibitem{b2.5a} T. Kajita, \Journal{\NPB\ (Proc. Suppl.)}{77}{123}{1999}.

\bibitem{b2.6} S. Fukuda {\it et al.} (The Super-Kamiokande Collaboration), \Journal{\PRL}{85}{3999}{2000}.

\bibitem{b2.7} J. Wilkerson, this Volume.

\end{thebibliography}
\end{document}